\documentclass[twocolumn,showpacs,amssymb]{revtex4}
\usepackage{indentfirst}
\usepackage{bm}
\usepackage{graphicx}
\usepackage{amsmath}
\begin{document}
\title{Pseudo-gap and vertex correction of electron-phonon interaction}
\author{Wei Fan\footnote{Corresponding author: Wei Fan, Email:fan@theory.issp.ac.cn; Project
supported by Knowledge Innovation Project of Chinese Academy of
Sciences.}}
\affiliation{ Key Laboratory of Materials Physics,
 Institute of Solid State Physics, Hefei Institutes of Physical
Sciences, Chinese Academy of Sciences, 230031-Hefei, People's Republic
of China}
\date{\today}

\begin{abstract}
The strong-coupling Eliashberg theory plus vertex correction is used
to calculate the maps of transition temperature (T$_{c}$) in
parameter-space characterizing superconductivity. Based on these
T$_{c}$ maps, complex crossover behaviors are found when
electron-phonon interaction increases from weak-coupling region to
strong-coupling region. The doping-dependent T$_{c}$ of cuprate
superconductors and most importantly the pseudo-gap can be explained
as the effects of vertex correction.
\end{abstract}
 \maketitle


The standard strong-coupling theory has no bound on T$_{c}$. Recently,
the Eliashberg functions $\alpha^{2}F(\omega)$ extracted from the
measurements of infrared optical conductivity~\cite{Heumen1} and ARPES
spectrum~\cite{Ruiz1} for copper-oxides superconductors predicted very
strong electron-phonon interaction and very high T$_{c}$ over the
experimental values~\cite{Heumen1}. The T$_{c}$ in mean-field
approximation of Eliashberg theory is higher than experimental
T$_{c}$. In the situation of very strong electron-phonon coupling,
non-adiabatic effects of electron-ion system will be so important that
the electrons are dressed heavily by lattice vibrations and the
conventional strong-coupling theory needs to be generalized to include
the non-adiabatic effects or the vertex corrects beyond Migdal's
theorem~\cite{Kostur1,Grimaldi1,Fan1}. The behavior of crossovers when
electron-phonon interaction evolving from weak-coupling region to
strong-coupling region were found in different theoretical
calculations~\cite{Paci1,Freericks1,Nasu1}. These crossovers are
expected to prevent T$_{c}$ from infinitely increasing with
electron-phonon interaction. If the electron-phonon interaction is the
underlying pairing-mechanism for unconventional superconductors such
as the cuprate superconductors, it should provide reasonable
explanations of pseudo-gap and the dome-shape of doping-dependent
T$_{c}$.

In previous works~\cite{Fan1,Fan2,Fan3}, we predicted that the highest
T$_{c}$ of cuprate superconductors is close to 160K consistent with
the present record of cuprate superconductors~\cite{Fan1} and
successfully explained the spatial anti-correlation between energy gap
and phonon energy for cuprate superconductor Bi2212~\cite{Fan2}. The
up-limit of T$_{c}$ for iron-based superconductors is about
90K~\cite{Fan1} and there are 34-35 K space left to increase T$_{c}$
beyond present record about 55-56 K~\cite{Ren1}. If the effects of
vertex correction (or non-adiabatic effects) are strong, it's very
difficult to realize home-temperature superconductor in high-pressure
metal hydrogen and other hydrogen-rich materials~\cite{Fan3}.

In this paper, the T$_{c}$ maps including the influences of vertex
corrections are studied. Complex crossovers are found on these T$_{c}$
maps when the parameter $\lambda$ of electron-phonon coupling
increases from weak-coupling region to strong-coupling region. These
crossovers are very close to the well known $\lambda$=2 at which the
value of T$_{c}$ reaches its maximum~\cite{McMillan1}.  The pseudo-gap
and phase diagram with same topology as the phase diagram of
doping-dependent T$_{c}$ for cuprate superconductor are explained as
the effects of vertex correction. The interplay of vertex correction
and Coulomb interaction can suppress theoretical T$_{c}$ to access
experimental values~\cite{Heumen1}

\begin{figure}[h]
\includegraphics[width=0.48\textwidth]{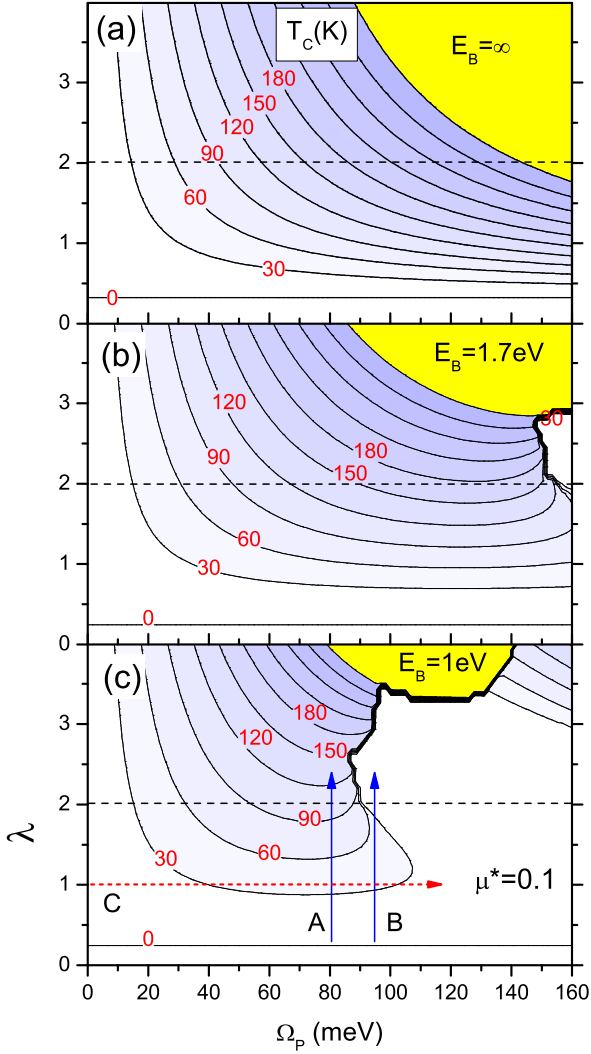}
\caption{\label{fig1} The evolution of T$_{c}$ map on
$\lambda$-$\Omega_{P}$ plane with increasing strengths of vertex
corrections (decreasing effective band-width E$_{B}$) with (a)
E$_{B}=\infty$, (b) E$_{B}$=1.7 eV and (c) E$_{B}$=1 eV. The Coulomb
pseudo-potential $\mu^{*}$=0.1.}
\end{figure}

The calculations of vertex corrections are greatly simplified under
isotropic approximation because the electron-phonon interactions are
included in the vertex corrections only by the functions of
electron-phonon interaction $\lambda_{n}$ defined as
$\lambda_{n}=2\int^{\infty}_{0}d\omega\alpha^{2}F(\omega)\omega/(\omega^{2}+\omega_{n}^{2})$.
When temperature is very close to T$_{c}$ the energy-gap
equation~\cite{Kostur1,Fan1} is simplified to
$\sum_{n'=-\infty}^{+\infty}K_{nn'}(\Delta_{n'}/|\omega_{n'}|)=0$. The
kernel matrix is expressed as
\begin{eqnarray}\label{GapKN}
 K_{nn'}&=&[\lambda_{n-n'}B_{nn'}-\mu^{*}+C_{nn'}]a_{n'}-\delta_{nn'}H_{n'},
 \\ \nonumber
 H_{n'}&=&\sum_{n''=-\infty}^{+\infty}[
 \frac{\delta_{n'n''}|\omega_{n''}|}{\pi
 k_{B}T}+\lambda_{n'-n''}A_{n'n''}s_{n'}s_{n''}a_{n''}].
 \end{eqnarray}
\noindent where the parameters $A_{nn'}=1-V^{A}_{nn'}$,
$B_{nn'}=1-V^{B}_{nn'}$, $s_{n}=\omega_{n}/|\omega_{n}|$ and
$a_{n}=(2/\pi)\arctan(E_{B}/Z_{n}|\omega_{n}|)$. The three parameters
of vertex correction $V^{A}_{nn'}$, $V^{B}_{nn'}$ and $C_{nn'}$ can be
found in Ref.\cite{Fan1}. The Coulomb pseudo-potential is defined as
$\mu^{*}=\mu_{0}/(1+\mu_{0}\ln(E_{B}/\omega_{0}))$, where
$\mu_{0}=N(0)U$, U the Coulomb interaction between electrons and
$\omega_{0}$ characteristic energy of typical phonon correlated to
superconductivity. If the vertex corrections are ignored, three
parameters $V^{A}_{nn'}$, $V^{B}_{nn'}$ and $C_{nn'}$ are all equal to
zero and the kernel Eq.(\ref{GapKN}) of energy-gap equation reduces to
the general form without vertex correction~\cite{Allen1} after some
symmetrizations and simplifications. It's convenient that the
$K_{nn'}$ matrix is symmetrized as in Ref.\cite{Allen1}. The
Eliashberg functions $\alpha^{2}F(\omega)$ have the same approximation
as in Ref.\cite{Scalapino1}. Other details in our calculations can be
found in Ref~\cite{Fan1,Fan3}

\begin{figure}[h]
\includegraphics[width=0.48\textwidth]{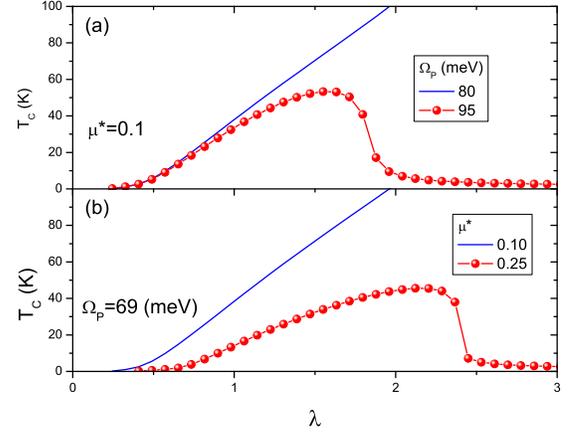}
\caption{\label{fig2} (a) The T$_{c}$ change along two arrows shown in
Fig.\ref{fig1}(c) with fixed phonon energies $\Omega_{P}$=80 meV and
95 meV respectively. (b) The T$_{c}$ change along two arrows shown in
Fig.\ref{fig3}(c) with fixed Coulomb pseudo-potentials $\mu^{*}$=0.10
and 0.25 respectively.}
\end{figure}

The parameter $\Omega_{P}/E_{B}$ measures and controls the magnitude
of vertex correction in perturbing calculation.  From electron point
of view, the vertex correction or non-adiabatic effect can be
controlled by the effective band-width E$_{B}$, on the other hand,
from ion point of view, it can be controlled by the cutoff
$\omega_{0}$ of phonon energy or $\Omega_{P}$ in Einstein model. In
this work, the vertex correction is controlled by the effective
band-width E$_{B}$ within the range from 0.5 eV to 5 eV. The situation
E$_{B}=\infty$ is equivalent to no vertex correction. The smaller
E$_{B}$ means possible the stronger vertex correction.

\begin{figure}[h]
\includegraphics[width=0.48\textwidth]{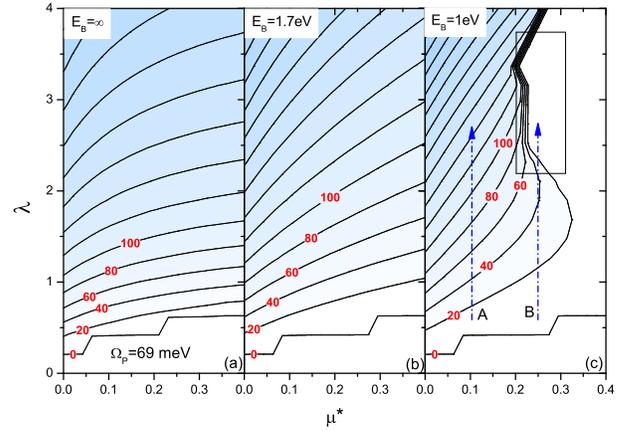}
\caption{\label{fig3}The evolutions of T$_{c}$ map on
$\mu^{*}$-$\lambda$ plane ($\Omega_{P}$=69 meV) with decreasing
effective band-width (a) E$_{B}=\infty$, (b) E$_{B}$=1.7 eV and (c)
E$_{B}$=1.0 eV. }
\end{figure}

The Fig.\ref{fig1}(a,b,c) illustrate the evolution of T$_{c}$ map on
$\lambda$-$\Omega_{P}$ plane with decreasing E$_{B}$. The
Fig.\ref{fig1}(a) is the T$_{c}$ map having been obtained in the
previous work without considering vertex corrections~\cite{Fan1}. When
E$_{B}$=1.7 eV, the large deformation of T$_{c}$ map with strong
vertex correction is shown in Fig.\ref{fig1}(b) near the well known
$\lambda$=2.0 in the region of high phonon energy. With E$_{B}$
decreasing to 1 eV further, the region with strong vertex correction
rapidly expands and occupies large part of parameter space with
$\Omega_{P}$$>$80 meV in Fig.\ref{fig1}(c). In the region
$\Omega_{P}$$<$80 meV, the T$_{c}$ is strongly suppressed however
there are no discontinuous changes of T$_{c}$ or breaking of contour
lines. An important result from the Fig.\ref{fig1} is that T$_{c}$
does not change with $\lambda$ monotonously if phonon energy
$\Omega_{P}$ is high enough. The Fig.\ref{fig2}(a) shows the changes
of T$_{c}$ with $\lambda$ along two arrows A and B shown in
Fig.\ref{fig1}(c). If $\Omega_{P}$=80 meV, the T$_{c}$ monotonously
increases with $\lambda$. However for $\Omega_{P}$=90 meV, the T$_{c}$
first increases with $\lambda$, reaches the maximum at $\lambda\sim
1.5-1.7$ and then quickly decreases with increasing $\lambda$. Further
increasing $\lambda$$>$2, T$_{c}$ will be very low due to strong
vertex corrections. The non-monotonous $\lambda$-dependent T$_{c}$ in
Fig.\ref{fig2}(a) had been found in the non-adiabatic theory of
superconductivity~\cite{Paci1}. Some crossover behaviors from weak
coupling to strong coupling region had been predicted in
Holstein-Hubbard model solved numerically by quantum Monte Carlo
method~\cite{Freericks1} and in polaron theory~\cite{Nasu1}. It's very
reasonable that the non-monotonous $\lambda$-dependent T$_{c}$ is
equivalent to the crossovers found in QMC calculation
~\cite{Freericks1} and polaron theory~\cite{Nasu1}. So only the
leading vertex correction can describe qualitatively very well the
electron-phonon interaction in strong coupling region.

\begin{figure}[h]
\includegraphics[width=0.48\textwidth]{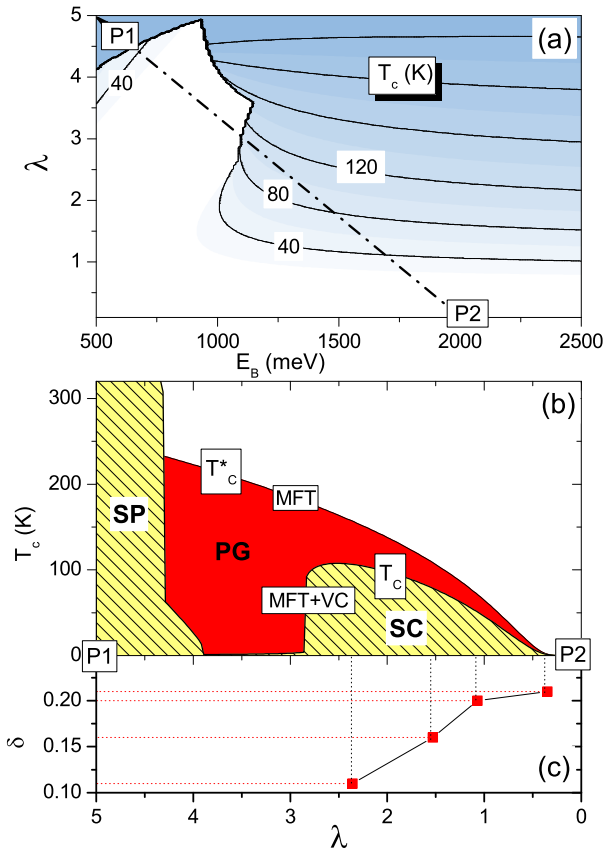}
\caption{\label{fig4}(a) The T$_{c}$ map on $E_{B}$-$\lambda$ plane
with $\mu^{*}$=0.25 and $\Omega_{P}$=72 meV. (b) The open circle line
is the evolution of T$_{c}$ from $P1$ to $P2$ in (a) but $\mu^{*}$
linearly decreases from 0.3 to 0.1. The solid line T$_{c}^{*}$ is the
standard results in strong coupling theory without vertex correction.
(c) The $\delta$-$\lambda$ relation is adopted in Ref.\cite{Heumen1}}
\end{figure}

The Fig.\ref{fig3}(a) is the normal T$_{c}$ map on $\mu^{*}$-$\lambda$
plane without vertex correction~\cite{Fan1}. The figure shows that
when $\mu^{*}$$>$0.2, T$_{c}$ is insensitive to the change of
$\mu^{*}$. The breaking contour lines with T$_{c}$=0 K are because of
the inaccurate calculations when T$_{c}$$<$0.1 K if only $N$=200
Matsubara energies are used. The contour lines with T$_{c}$$>$0.1 K
are accurate enough. If the Coulomb pseudo-potential and vertex
correction work together, the situation will change drastically and
some new interesting results will appear. The large deformations are
found in Fig.\ref{fig3}(c) if E$_{B}$ decreases to 1.0 eV. As
expected, the large deformations and discontinuous changes of contour
lines appear on the T$_{c}$ map when $\mu^{*}$$>$0.20. The contour
lines with iso-values from T$_{c}$=20 K to 200 K are packed together
within the rectangle region in Fig.\ref{fig3}(c) with
0.15$<$$\mu^{*}$$<$0.25 and $\lambda$$>$2. The figure clearly shows
that if the Coulomb pseudo-potential $\mu^{*}$ is larger enough, the
T$_{c}$ will change with $\lambda$ non-monotonously.  The changes of
T$_{c}$ along two arrows with $\mu^{*}$=0.1 and 0.25 are plotted in
Fig.\ref{fig2}(b). For $\mu^{*}$=0.25, T$_{c}$ first increases with
$\lambda$ until reaches the maximum at $\lambda$=2.2 and then sharply
decreases to smaller value at $\lambda$=2.5. The crossover behavior is
enhanced by strong Coulomb interaction.

\begin{figure}[h]
\includegraphics[width=0.48\textwidth]{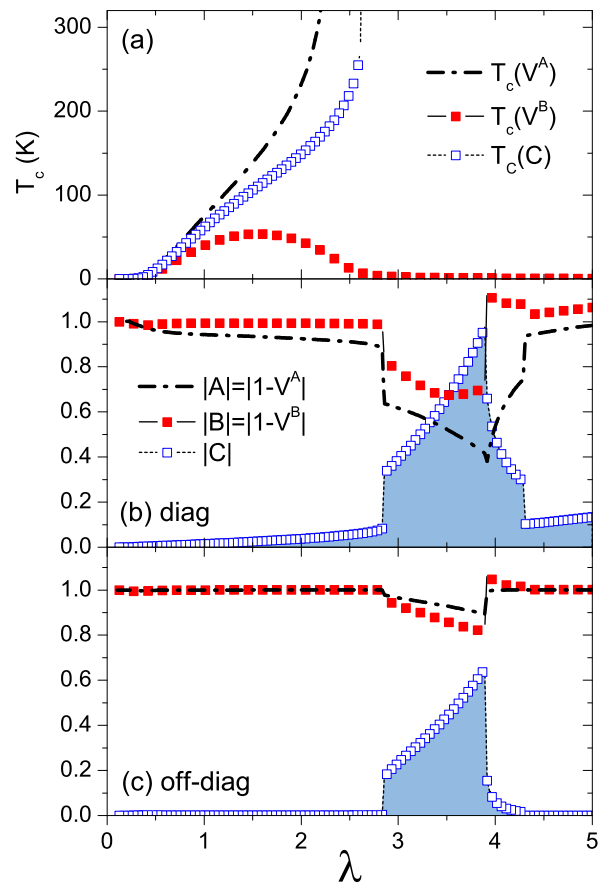}
\caption{\label{fig5}(a) The effects of three non-adiabatic parameters
$V^{A}_{nn'}$,$V^{B}_{nn'}$ and $C_{nn'}$ on T$_{c}$. (b,c) The
changes of average values of diagonal and off-diagonal matrix elements
of $|A_{nn'}|=|1-V^{A}_{nn'}|$, $|B_{nn'}|=|1-V^{B}_{nn'}|$ and
$|C_{nn'}|$ with increasing $\lambda$.}
\end{figure}

The T$_{c}$ map on E$_{B}$-$\lambda$ plane is presented in
Fig.\ref{fig4}(a) with $\Omega_{P}$=72 meV. If E$_{B}$ increases but
$\lambda$ keeps unchanged, the T$_{c}$ monotonously increases with
E$_{B}$ until to the limit of non-vertex-correction. More
interestingly, on this map, the T$_{c}$ is non-monotonous dependent on
E$_{B}$ along straight line from $P1$ to $P2$ companying by the
decrease of $\lambda$ from 5.0 to 0.2. The non-monotonous dependence
of T$_{c}$ on effective band-width E$_{B}$ is equivalent to the
band-filling effect of T$_{c}$. Our results show that, if
$\Omega_{P}>$80 meV, the suppression of T$_{C}$ will be more
prominently.

The values of T$_{c}$ obtained from standard strong-coupling theory
are generally higher than those measured in experiments. The
copper-oxides superconductors Bi$_{2}$Sr$_{2}$CaCu$_{2}$O$_{8+\delta}$
and Bi$_{2}$Sr$_{2}$Cu$_{2}$O$_{6+\delta}$ studied in Ref
~\cite{Heumen1,Ruiz1} have very strong electron-phonon interactions
$\lambda$$\sim$2.36-2.85 and overestimated T$_{c}$ in underdoped
samples. With increasing doping $\delta$, the values of $\lambda$
decrease to about 0.35-1.42~\cite{Heumen1}. The effective band-widths
E$_{B}$ of conducting electrons for these cuprates are distributed
from 1 eV to 3 eV. The effective phonon energies are distributed from
50 meV to 80 meV. We re-calculate the values of T$_{c}$ along straight
line $P1$-$P2$ in Fig.\ref{fig4}(a) under assumption that the Coulomb
interaction is strong in underdoped region $\mu^{*}$=0.3 at $P1$ and
weak in overdoped region $\mu^{*}$=0.1 at $P2$. For simplicity,
$\mu^{*}$ linearly decreases from 0.3 at $P1$ to 0.1 at $P2$. As shown
in Fig.\ref{fig4}(b), if $\lambda<$4.0, the values of T$_{c}$ are
reduced from around 200 K to lower than 150 K and close to
experimental values~\cite{Heumen1}. In strong-coupling region
4.0$>\lambda>$3.0, T$_{c}$ is very low. Our results are provided an
explanation to pseudo-gap in underdoped region shown in
Fig.\ref{fig4}(b). The cooper-pairs pre-form at T$_{c}^{*}$ the
transition temperature of mean field theory (MFT) or the standard
strong-coupling theory. However strong non-adiabatic effects induce
the instability of cooper-pairs and the real T$_{c}$ has lower value.
The T$^{*}_{c}$ degenerating with T$_{c}$ in overdoped region is
similar to the example (1) of Fig.9 in Ref.\cite{Norman1}. The
pre-formed cooper-pairs in cuprate superconductors are supported by
measurements of Nernst effect~\cite{Xu1}, specific-heat~\cite{Wen1}
and many other methods.

An interesting result is that at very strong coupling $\lambda>$4.0,
the effects of vertex corrections superficially become weak. Even
there are positive vertex correction that had been found in other
work~\cite{Paci1}. The electronic states in region $SP$ with
$\lambda>$4.0 are strong-coupling pairs~\cite{Chakraverty1}. The
Fig.\ref{fig4}(b) shows a crossover from BCS state to strong-coupling
pairs state with increasing electron-phonon interaction $\lambda$.
It's obviously that Fig.\ref{fig4}(b) has the same topology as the
well-known T$_{c}$-doping phase diagram. It's dependent on whether the
parameter $\lambda$ electron-phonon interaction decreases with
increasing doping or not. This point had been proved in recent
experiments~\cite{Heumen1,Ruiz1}. The $\delta-\lambda$ curve in
Fig.\ref{fig4}(c) is based on data in Ref.\cite{Heumen1}. It's urged
that there will be other experiments supporting this point.

In order to analyze our results more deeply, we present individually
the effects of non-adiabatic parameters $V^{A}_{nn'}$, $V^{B}_{nn'}$
and $C_{nn'}$ on T$_{c}$ in Fig.\ref{fig5}(a). The T$_{c}$-$\lambda$
curve labeled with $V^{B}$ is calculated by allowing $V^{B}_{nn'}\ne
0$ and setting $V^{A}_{nn'}$=0 and $C_{nn'}$=0. Other curves are
obtained with the same manner. We can find that the dome shape curve
of T$_{c}$ in the region $\lambda<$4.0 is generated by the effects of
$V^{B}_{nn'}$. In the region $\lambda>$4.0, the effects of
$V^{A}_{nn'}$ and $C_{nn'}$ cancel the effects of $V^{B}_{nn'}$ so
that the strong coupling pairs in mean field approximation are
restored and even have higher T$_{c}$. This fact can be clarified from
the Fig.\ref{fig5}(b,c) in that the averages of  the absolute values
of diagonal and off-diagonal elements of parameter matrix
$|A_{nn'}|=|1-V^{A}_{nn'}|$, $|B_{nn'}|=|1-V_{nn'}|$ and $|C_{nn'}|$
have larger changes mainly in the region 2.5$<\lambda<$4.0. Moreover
in the region $\lambda>$4.3, the $|A_{nn'}|$ and $|B_{nn'}|$ are close
to normal values 1.0 just as in weak coupling region. Additionally,
the average values of diagonal elements of parameter matrix
$|C_{nn'}|$ and $|B_{nn'}|$ steadily increase with $\lambda$ and lead
to positive vertex-correction.

In summary the non-monotonous changes of T$_{c}$ with increasing
$\lambda$ show the crossover behaviors near $\lambda$=2 when $\lambda$
evolving from weak-coupling region to strong-coupling region. The
crossovers can explain both the pseudo-gap phenomenon and the dome
shape of doping dependent T$_{c}$ of cuprate superconductors. The
T$_{c}$ maps in the previous paper~\cite{Fan1} and the maps with
vertex corrections in this paper provide very comprehensive
understanding of superconductivity of superconductors. The author
thanks Prof. E. Cappelluti for very helpful discussions.

\end{document}